\documentclass[aps,prl,preprint,superscriptaddress,amsmath,amssymb,showpacs]{revtex4}

\usepackage{graphicx}
\usepackage{citesort}

\def\e{\mathrm{e}}
\def\iunit{\mathrm{i}}

\def\nn{\nonumber}

\def\sgn{\mathrm{sgn}}
\renewcommand{\leq}{\leqslant}

\begin{document}


\title{Exact scattering eigenstates, many-body bound states,\\[2pt]
and nonequilibrium current of an open quantum dot system\\
}


\author{Akinori Nishino}

\author{Takashi Imamura}

\author{Naomichi Hatano}

\affiliation{Institute of Industrial Science, The University of Tokyo,
4--6--1 Komaba, Meguro-ku, Tokyo, 153--8505}


\date{\today}

\begin{abstract}
We obtain an exact many-body scattering eigenstate 
in an open quantum dot system.
The scattering state is not in the form of the Bethe eigenstate
in the sense that the wave-number set of the incoming plane wave is not 
conserved during the scattering and many-body bound states appear.
By using the scattering state, we study 
the average nonequilibrium current 
through the quantum dot under a finite bias voltage.
The current-voltage characteristics that we obtained 
by taking the two-body bound state into account
is qualitatively similar to several known results.
\end{abstract}



\maketitle


Mesoscopic transport of interacting electrons
has attracted much interest recently~\cite{%
Ralph-Buhrman_92PRL-94PRL,%
GoldhaberGordon_98Nature,Cronenwett_98Science,%
Wiel-Franceschi-Fujisawa-Elzerman-Tarucha-Kouwenhoven_00Science}.
A remarkable feature of the mesoscopic system is 
the coherence length greater than the sample size. 
In the standard theory, the electron in the sample 
is described by the quantum mechanics
and dissipation is considered to occur only 
in reservoirs connected to the sample.
A well-known approach to the electric current
across the sample under a finite bias voltage is the Landauer formula, 
although the original one is restricted to the non-interacting case.
The Green's function is also employed to 
study the transport property~\cite{%
Rammer-Smith_86RMP,Meir-Wingreen-Lee_91PRL,Meir-Wingreen_92PRL,%
Hershfield-Davies-Wilkins_91PRL,%
Yeyati-MartinRodero-Flores_93PRL,Fujii-Ueda_03PRB}.
To discuss the effect of interactions in this framework, 
however, we would have to resort to 
a perturbation technique, which is generally a hard task.

In this Letter, we present an exact many-body scattering eigenstate
in an open quantum dot system and apply the eigenstate
to analysis of the nonequilibrium current.
The system we study is an open interacting resonant-level model (IRLM),
which consists of two leads of non-interacting spinless electrons 
that interact with an electron on a quantum dot in between the two leads.
Each lead is connected to a large reservoir.
First, we explicitly construct two- and three-electron scattering states,
which are free-electronic plane waves 
before scattering and, at the quantum dot, are partially scattered 
to a many-body bound state due to the Coulomb interaction.
Second, by using the scattering states, 
we calculate the quantum-mechanical expectation value of 
the current through the quantum dot
in the second order of the inverse system length.
Third, we study the statistical average of the nonequilibrium current
for a given finite bias voltage under the assumption 
that electrons are completely thermalized in each reservoir
before returning to the lead.

\begin{figure}
\includegraphics[width=90mm,clip]{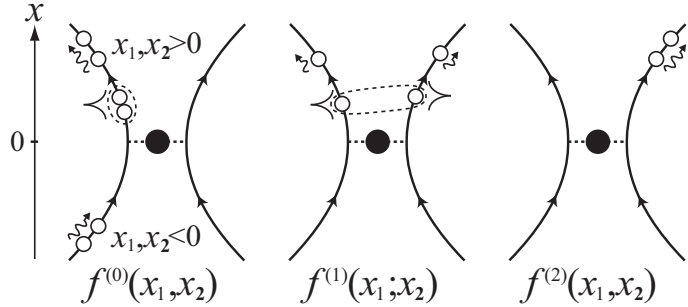}
\caption{\label{fig:bound-state} 
A two-electron scattering state
which contains incoming plane waves only in the left lead.
}
\end{figure}

Our study of the nonequilibrium current with 
scattering states is a genuine extension of the Landauer formula.
Our scattering states of the {\it open} system
are suitable for describing incident electrons 
thermalized to a free-electron state in each reservoir.
Some used the Bethe 
ansatz~\cite{Andrei_80PRL,Wiegmann_80PLA,Filyov-Wiegmann_80PLA}
to study the transport properties
of quantum dot systems~\cite{%
Konik-Saleur-Ludwig_01PRL-02PRB},
where the Landauer formula was formally applied to
the quasi-particles in a {\it closed} system in equilibrium.
However, the periodic boundary conditions 
imposed on the Bethe state are clearly different from 
the conditions adopted for the Landauer formula,
the conditions that the incident electrons are asymptotically free.
Recently, there have been a few attempts to study 
the transport properties with a scattering state
in the framework of the Lippmann-Schwinger (LS) equation~\cite{%
Lebedev-Lesovik-Blatter_08PRL,Dhar-Sen-Roy_08PRL}.
Our scattering state is shown a solution of the LS equation
associated with the open IRLM. 

A remarkable point of our solution is the appearance of 
{\it a many-body bound state} in the scattering eigenstate.
Another many-body bound state given by the Bethe ansatz
method is known to be the ground state of the Anderson model
in equilibrium~\cite{Kawakami-Okiji_81PLA-82JPSJ}.
Our bound state, on the other hand, is generated as a result of
the scattering of an incident free-electronic plane-wave state 
(Fig.~\ref{fig:bound-state}).
The interaction around the quantum dot is 
a necessary condition of the appearance of the bound state.
The nonequilibrium current is indeed affected 
by the interaction through the bound state.

The open IRLM out of equilibrium
has been studied with various 
approaches~\cite{Mehta-Andrei_06PRL,Doyon_07PRL,Golub_07PRB,%
Boulat-Saleur-Schmitteckert_08PRL,Boulat-Saleur_08PRB}.
We express the quantum-mechanical expectation value of 
the current as a series of the inverse system length
to consider the average current,
while the perturbative result~\cite{Golub_07PRB} gives
the average current as a series of the interaction parameter.
The qualitative behaviour of the current-voltage characteristics
that we obtain is similar to the results in Refs.~\cite{Doyon_07PRL,%
Golub_07PRB,Boulat-Saleur-Schmitteckert_08PRL}.
We remark that, in our results, the effect of the interaction 
appears in the quantum-mechanical expectation value, 
which differs from the result in Ref.~\cite{Mehta-Andrei_06PRL}.

The Hamiltonian of the open IRLM is given by 
\begin{align}
\label{eq:IRLM_12}
&H\!=\!\sum_{\alpha}\hspace{-2pt}\Big(
 \int_{-L/2}^{L/2}
 \hspace{-15pt}dx\;
 c^{\dagger}_{\alpha}(x)
 \frac{1}{\iunit }\frac{d}{dx}c_{\alpha}(x)
 \!+\!\Bar{t}
 \big(c^{\dagger}_{\alpha}(0)d\!+\!d^{\dagger}c_{\alpha}(0)\big)
 \Big)
 \nn\\
&\qquad
 +\epsilon_{d}d^{\dagger}d
 +\sum_{\alpha}
 Uc^{\dagger}_{\alpha}(0)c_{\alpha}(0)d^{\dagger}d,
\end{align}
where $c^{\dagger}_{\alpha}(x)$ and $c_{\alpha}(x)$
are creation- and annihilation-operators of the electrons
in the lead $\alpha(=1,2)$,
$d^{\dagger}$ and $d$ are those in the quantum dot,
$\Bar{t}=t/\sqrt{2}$ is the transfer integral 
between each lead and the dot, 
$\epsilon_{d}$ is the gate energy of the dot 
and $U(>0)$ expresses the Coulomb repulsion.
The dispersion relation in the leads is linearized
in the vicinity of the Fermi energy to be 
$E=v_{\mathrm{F}}k$, 
under the assumption that $t$, $\epsilon_{d}$ 
and $U$ are small compared with the Fermi energy~\cite{%
Andrei_80PRL,Wiegmann_80PLA,Filyov-Wiegmann_80PLA}.
For simplicity, we have set $v_{\mathrm{F}}=1$ 
in Eq.~\eqref{eq:IRLM_12}.
We treat the system as an open system in the limit $L\to\infty$.
The lead $\alpha$ is connected infinitely far way to a large reservoir
characterized by the Fermi distribution 
with the chemical potential $\mu_{\alpha}$.
Our goal is to calculate the statistical average
of the current
\begin{align}
\label{eq:current-op}
 I
&=\iunit \bar{t}\sum_{\alpha}(-)^{\alpha+1} 
 \big(c_{\alpha}^{\dagger}(0)d-d^{\dagger}c_{\alpha}(0)\big)
\end{align}
for the system under a finite bias voltage, $\mu_{1}>\mu_{2}$.

We consider the general form of eigenstates.
After the transformation
$c_{1/2}(x)=\big(c_{e}(x)\pm c_{o}(x)\big)/\sqrt{2}$,
the Hamiltonian~\eqref{eq:IRLM_12} is decomposed into
the even and odd parts. 
Due to the relations $[H, N_{e}+N_{d}]=[H, N_{o}]=0$
for the number operators 
$N_{e/o}=\int\! dx\, c_{e/o}^{\dagger}(x)c_{e/o}(x)$
and $N_{d}=d^{\dagger}d$,
the set $\{N_{e}+N_{d},N_{o}\}$
gives a good quantum number.
The $N$-electron state $|N,n\rangle$ 
in the sector with $N_{o}=n$
is expressed in the form
\begin{align}
\label{eq:general-N-state}
|N,n\rangle=
&\Big(\int\!\!dx^{N-n}dy^{n}
 g^{(n)}(x;y)
 c^{\dagger}_{e}(x_{1})\cdots c^{\dagger}_{e}(x_{N-n})
 c^{\dagger}_{o}(y_{1})\cdots c^{\dagger}_{o}(y_{n})
 \nn\\
&\!+\!\int\!\! dx^{N-n-1}dy^{n}
 e^{(n)}(x;y)
 c^{\dagger}_{e}(x_{1})\cdots c^{\dagger}_{e}(x_{N-n-1})d^{\dagger}
 c^{\dagger}_{o}(y_{1})\cdots c^{\dagger}_{o}(y_{n})
 \Big)|0\rangle,
\end{align}
where we put $e^{(N)}(x;y)=0$.
The functions $g^{(n)}(x;y)$ and $e^{(n)}(x;y)$
are antisymmetric with respect to the variables $\{x_{i}\}$ 
and with respect to $\{y_{i}\}$.
The one-electron eigenstate $|1,n;k\rangle$, $(n=0,1)$ 
with the energy eigenvalue $E=k$ is obtained
by inserting the eigenfunctions 
$g^{(0)}(x)=g_{k}(x)$, $e^{(0)}=e_{k}$ or $g^{(1)}(y)=h_{k}(y)$
into the general form~\eqref{eq:general-N-state}, where
\begin{align}
&g_{k}(x)=
 \frac{1}{\sqrt{2\pi}}\e^{\iunit kx}
 \big(\theta(-x)+\frac{e_{k}}{e_{k}^{\ast}}\theta(x)\big), 
 \nn\\
&h_{k}(x)=
 \frac{1}{\sqrt{2\pi}}\e^{\iunit kx},\quad
 e_{k}=
 \frac{1}{\sqrt{2\pi}}
 \frac{t}{k-\epsilon_{d}+\iunit \Bar{t}^{2}},
\end{align}
with the step function $\theta(x)$.
The linear combination
$|k\rangle=(|1,0;k\rangle+|1,1;k\rangle)/\sqrt{2}$ gives
a scattering state containing
an incoming electron only in the lead 1.
If we imposed periodic boundary conditions
to the leads, the wave number $k$ allowed 
for the eigenfunction $g^{(1)}(x)$ would be different from 
that for $g^{(0)}(x)$.
Thus, even in the non-interacting case, the scattering state
$|k\rangle$ is inconsistent with the periodic boundary conditions.

For $N=2$,
the eigenvalue problem $H|2,n\rangle=E|2,n\rangle$
is cast into a set of the Schr\"odinger equations:
\begin{align}
&\Big(\frac{1}{\iunit }(\partial_{1}\!+\!\partial_{2})\!-\!E\Big)
 g^{(0)}(x_{1},x_{2})
 \!-\!\frac{t}{2}
 \big(\delta(x_{1})e^{(0)}(x_{2})\!-\! e^{(0)}(x_{1})\delta(x_{2})\big)
 \!=\! 0,
 \nn\\
&\Big(\frac{1}{\iunit }\frac{d}{dx}
 \!+\!U\delta(x)\!+\!\epsilon_{d}\!-\!E\Big)e^{(0)}(x) 
 \!+\! 2t g^{(0)}(x,0)=0,
 \nn\\
&\Big(\frac{1}{\iunit }(\partial_{1}\!+\!\partial_{2})\!-\!E\Big)
 g^{(1)}(x_{1};x_{2})
 \!+\! t\delta(x_{1})e^{(1)}(x_{2})\!=\! 0,
 \nn\\
&\Big(\frac{1}{\iunit }\frac{d}{dx}
 \!+\!U\delta(x)\!+\!\epsilon_{d}\!-\!E\Big)e^{(1)}(x) 
 \!+\! t g^{(1)}(0;x)=0,
 \nn\\
&\Big(\frac{1}{\iunit }(\partial_{1}\!+\!\partial_{2})\!-\!E\Big)
 g^{(2)}(x_{1},x_{2})
 \!=\! 0.
 \label{eq:Sch-eq_2-state_IRLM}
\end{align}
We construct the eigenfunctions $g^{(0)}(x_{1},x_{2})$,
$g^{(1)}(x_{1};x_{2})$ and $g^{(2)}(x_{1},x_{2})$ 
by imposing the conditions that, 
in the region $x_{1}, x_{2}<0$, they are free-electronic plane waves.
The eigenfunction $g^{(0)}(x_{1},x_{2})$ 
is discontinuous at $x_{1}=0$ and $x_{2}=0$,
$g^{(1)}(x_{1};x_{2})$ at $x_{1}=0$,
and $e^{(0,1)}(x)$ at $x=0$.
The value of the functions at the discontinuous point
cannot be determined by Eqs.~\eqref{eq:Sch-eq_2-state_IRLM}.
We then set $g^{(0)}(x,0)=(g^{(0)}(x,0+)+g^{(0)}(x,0-))/2$ 
and so on.
The function $g^{(2)}(x_{1},x_{2})$ 
should be a free-electron eigenfunction.
The eigenfunctions with the energy eigenvalue $E=k_{1}+k_{2}$,
($k_{1}, k_{2}\in\mathbb{R}$) are then given as follows:
\begin{align}
\label{eq:2-eigenfunction}
&2g^{(0)}(x_{1},x_{2})
 \!=\sum_{Q}\sgn(Q)
 \big(g_{k_{1}}(x_{Q_{1}})g_{k_{2}}(x_{Q_{2}})
 \!+\! u Z_{12}(x_{Q_{1}Q_{2}})
  \e^{\iunit Ex_{Q_{2}}}\theta(x_{Q_{1}})
  \big),  
 \nn\\
&e^{(0)}(x)
 \!=\! g_{k_{1}}(x)e_{k_{2}}\!-\! g_{k_{2}}(x)e_{k_{1}}
 \!+\!\frac{u}{\iunit t}Z_{12}(-x)
  \e^{\iunit Ex}, 
 \nn\\
&g^{(1)}(x_{1};x_{2})
 \!=\! g_{k_{1}}(x_{1})h_{k_{2}}(x_{2})
 \!-\! u X_{1}(x_{12})
  \e^{\iunit Ex_{2}}\theta(x_{1}), 
 \nn\\
&e^{(1)}(x)
 \!=\! e_{k_{1}}h_{k_{2}}(x)
 \!+\!\frac{u}{\iunit t}X_{1}(-x)
  \e^{\iunit Ex},
 \nn\\
&2g^{(2)}(x_{1},x_{2})
 \!=\sum_{Q}
  \sgn(Q)h_{k_{1}}(x_{Q_{1}})h_{k_{2}}(x_{Q_{2}}),
\end{align}
where $Q=(Q_{1},Q_{2})$ is a permutation of $(1,2)$,
$x_{ij}=x_{i}-x_{j}$, $u=2U/(2+\iunit U)$ and
\begin{align}
&Z_{ij}(x)=(k_{i}-k_{j})e_{k_{i}}e_{k_{j}}
 \e^{\iunit (\epsilon_{d}-\iunit\Bar{t}^{2})x}\theta(-x),
 \nn\\ 
&X_{i}(x)=\frac{t}{\sqrt{2\pi}}e_{k_{i}}
 \e^{\iunit(\epsilon_{d}-\iunit\Bar{t}^{2})x}\theta(-x).
\end{align}
The wave-number set $\{k_{1},k_{2}\}$ in each of the eigenfunctions 
$g^{(0)}(x_{1},x_{2})$ and $g^{(1)}(x_{1};x_{2})$ is 
not conserved during the scattering; 
the plane wave with $\{k_{1}, k_{2}\}$ is partially scattered to 
that with $\{\epsilon_{d}-\iunit\Bar{t}^{2},
E-\epsilon_{d}+\iunit\Bar{t}^{2}\}$ in the region $x_{1}, x_{2}>0$.
In this sense, they are not the Bethe 
eigenfunctions~\cite{Filyov-Wiegmann_80PLA,%
Mehta-Andrei_06PRL,Nishino-Hatano_07JPSJ}.
We have found similar eigenfunctions
in the Anderson model~\cite{Imamura-Nishino-Hatano_09preprint}.

The second term of each of the first four 
eigenfunctions~\eqref{eq:2-eigenfunction} 
comes from the Coulomb interaction.
The imaginary part of the wave numbers, $\iunit\Bar{t}^{2}$,
indicates the appearance of {\it a two-body bound state}
$\e^{-\Bar{t}^{2}|x_{12}|}$.
The interaction is a necessary condition
of the appearance of the bound state 
and the strength of binding is determined by 
the transfer integral $\Bar{t}$.
A similar two-photon bound state 
has been found in a one-dimensional waveguide 
coupled to a two-level system~\cite{Shen-Fan_07PRL},
where the bound state has been obtained 
through an ``S-matrix" acting on the Hilbert space 
of free two photons and the eigenstate including 
the bound state has not been constructed.

We obtain two-electron eigenstates 
by inserting the eigenfunctions~\eqref{eq:2-eigenfunction} 
into the form~\eqref{eq:general-N-state};
we denote them by $|2,n;k_{1},k_{2}\rangle$, ($n=0,1,2$).
We notice that, by exchanging $k_{1}$ and $k_{2}$ in 
$|2,1;k_{1},k_{2}\rangle$, 
we have another eigenstate $|2,1;k_{2},k_{1}\rangle$
with the same energy.
The four eigenstates satisfy the orthonormal relations
in the limit $L\to\infty$:
\begin{align}
&\langle 2,n;k_{1},k_{2}|2,n;k_{1}^{\prime},k_{2}^{\prime}\rangle
 =\delta(k_{1}\!-\!k_{1}^{\prime})
  \delta(k_{2}\!-\!k_{2}^{\prime})
  \!-\!
  \delta(k_{1}\!-\!k_{2}^{\prime})
  \delta(k_{2}\!-\!k_{1}^{\prime}),
 \quad (n=0,2)
 \nn\\
&\langle 2,1;k_{1},k_{2}|2,1;k_{1}^{\prime},k_{2}^{\prime}\rangle
 =\delta(k_{1}\!-\!k_{1}^{\prime})
  \delta(k_{2}\!-\!k_{2}^{\prime}).
\end{align}

In principle,
we can construct eigenstates for a few electrons.
For example, the three-electron eigenfunctions
in the sector with $N_{o}=0$ are given by
\begin{align}
&3! g^{(0)}(x_{1},x_{2},x_{3})
 =\sum_{P}\sgn(P)
 g_{k_{P_{1}}}(x_{1})g_{k_{P_{2}}}(x_{2})g_{k_{P_{3}}}(x_{3})
 \nn\\
&\!+\!\frac{u}{2}\sum_{P,Q}\sgn(PQ)
 g_{k_{P_{1}}}(x_{Q_{1}})Z_{P_{2}P_{3}}(x_{Q_{2}Q_{3}})
 \e^{\iunit(k_{P_{2}}+k_{P_{3}})x_{Q_{3}}}
 \theta(x_{Q_{2}})
 \nn\\
&\!-\!\frac{u^{2}}{2\iunit}
 \sum_{P,Q}\sgn(PQ)
 h_{k_{P_{1}}}(x_{Q_{2}})Z_{P_{2}P_{3}}(x_{Q_{1}Q_{3}})
 \e^{\iunit(k_{P_{2}}+k_{P_{3}})x_{Q_{3}}}
 \theta(x_{Q_{3}Q_{2}})\theta(x_{Q_{2}Q_{1}})\theta(x_{Q_{1}}),
 \nn\\
&2! e^{(0)}(x_{1},x_{2})
 =\sum_{P}\sgn(P)
  g_{k_{P_{1}}}(x_{1})g_{k_{P_{2}}}(x_{2})e_{k_{P_{3}}}
 \nn\\
&\!+\!\frac{u}{2\iunit t}\sum_{P,R}\sgn(PR)
  g_{k_{P_{1}}}(x_{R_{1}})Z_{P_{2}P_{3}}(-x_{R_{2}})
 \e^{\iunit(k_{P_{2}}+k_{P_{3}})x_{R_{2}}}
 \nn\\
&\!+\!\frac{u}{2}
 \sum_{P,R}\sgn(PR)
 Z_{P_{2}P_{3}}(x_{R_{1}R_{2}})
 \e^{\iunit(k_{P_{2}}+k_{P_{3}})x_{R_{2}}}
 e_{k_{P_{1}}}\theta(x_{R_{1}})
 \nn\\
&\!-\!\frac{u^{2}}{2t}
 \sum_{P,R}\sgn(PR)
 h_{k_{P_{1}}}(x_{R_{1}})Z_{P_{2}P_{3}}(-x_{R_{2}})
 \e^{\iunit(k_{P_{2}}+k_{P_{3}})x_{R_{2}}}
 \theta(x_{R_{2}R_{1}})\theta(x_{R_{1}}).
\end{align}
Here $P=(P_{1},P_{2},P_{3})$ and 
$Q=(Q_{1},Q_{2},Q_{3})$ are permutations of $(1,2,3)$
and $R=(R_{1},R_{2})$ is 
that of $(1,2)$.
The third term of the eigenfunction
$g^{(0)}(x_{1},x_{2},x_{3})$ indicates 
a new three-body bound state.
The eigenstates in other sectors 
with $N_{o}=1,2,3$ are constructed in similar ways.

Now we construct a scattering eigenstate
by taking a linear combination of 
the four two-electron eigenstates as
\begin{align}
&|k_{1},k_{2}\rangle
 =A|2,0;k_{1},k_{2}\rangle
 +B_{1}|2,1;k_{1},k_{2}\rangle
 -B_{2}|2,1;k_{2},k_{1}\rangle
 +C|2,2;k_{1},k_{2}\rangle.
\label{eq:linear-comb}
\end{align}
Going from the eigenfunctions in terms of the even and odd parts
back to the ones in terms of the leads 1 and 2,
we have $f^{(0/2)}(x_{1},x_{2})
=\langle 0|c_{1/2}(x_{2})c_{1/2}(x_{1})|k_{1},k_{2}\rangle$
and $f^{(1)}(x_{1};x_{2})
=\langle 0|c_{2}(x_{2})c_{1}(x_{1})|k_{1},k_{2}\rangle$.
By choosing $A=B_{1}=B_{2}=C=1/2$ in Eq.~\eqref{eq:linear-comb},
we obtain the scattering state which contains
an incoming two-electron plane wave only in the lead 1, i.e.,
$f^{(1)}(x_{1};x_{2})=f^{(2)}(x_{1},x_{2})=0$ for $x_{1}, x_{2}<0$, 
which is depicted in Fig.~\ref{fig:bound-state}.
In the same way, by choosing 
$A=-B_{1}=B_{2}=-C=1/2$,
we obtain the scattering state which contains
an incoming one-electron plane wave in each lead, i.e.,
$f^{(0,2)}(x_{1},x_{2})=0$ for $x_{1}, x_{2}<0$.
We denote the former/latter scattering state by 
$|k_{1},k_{2}\rangle_{\pm}$.
Each scattering state is shown to be a solution of the LS equation 
whose incident state is a free-electron plane-wave state,
where the incident state means an eigenstate
of the Hamiltonian~\eqref{eq:IRLM_12} with $\Bar{t}=0$.
On the other hand, the scattering state 
constructed from the Bethe 
eigenstates~\cite{Mehta-Andrei_06PRL,Nishino-Hatano_07JPSJ}
is interpreted as the solution associated with 
an incident state that depends on the parameter $U$.
We remark that the scattering states are also
constructed from a superposition of an infinite number of
the degenerate Bethe eigenstates~\cite{Nishino-Hatano_07JPSJ}.

We use the two-electron scattering states 
to calculate the quantum-mechanical expectation 
value of the current $I$ in Eq.~\eqref{eq:current-op}.
The expectation value with respect to 
the scattering state $|k_{1},k_{2}\rangle_{\pm}$,
$(k_{1}<k_{2})$ is calculated as 
\begin{align}
&\frac{\langle k_{1},k_{2}|I|k_{1},k_{2}\rangle_{\pm}}
 {\langle k_{1},k_{2}|k_{1},k_{2}\rangle_{\pm}}
 \!=\!\frac{2\pi}{L}\big(I_{0}(k_{1})\!\pm\! I_{0}(k_{2})\big)
 \!+\!\frac{4\pi^{2}}{L^{2}}I_{\pm}(k_{1},k_{2}),
 \nn\\
&I_{0}(k)
 \!=\!-\frac{t}{\sqrt{2\pi}}\mathrm{Im}(e_{k}),
 \nn\\
&I_{\pm}(k,\!h)
 \!=\!\frac{k\!-\!h}{t\sqrt{2\pi}}
 \big(\mathrm{Re}(e_{h})\mathrm{Im}(ue_{k}^{2})
 \!\pm\!\mathrm{Re}(e_{k})\mathrm{Im}(ue_{h}^{2})\big),
\label{eq:current-QM}
\end{align}
where $L=2\pi\delta(0)$ is the length of the system.
The first term of order $L^{-1}$
gives the current of non-interacting electrons. 
The correction term of order $L^{-2}$ containing 
$I_{\pm}(k_{1},k_{2})$
is due to the two-body bound state.

\begin{figure}[t]
\includegraphics[width=90mm,clip]{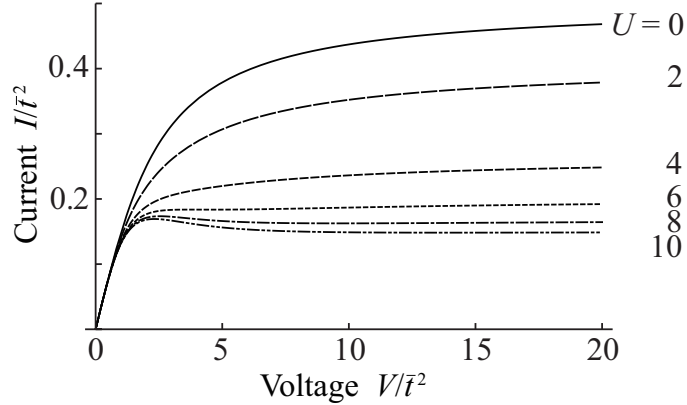}
\caption{\label{fig:current} 
A current-voltage characteristics of the average 
current for $\epsilon_{d}=0$ 
and $U=0$, $2$, $4$, $6$, $8$, $10$.}
\end{figure}

We find that, in the limit $L, N\to\infty$,
the correction term in Eqs.~\eqref{eq:current-QM}
contributes to the current.
In the spirit of the Landauer formula, we assume that 
electrons are completely thermalized in each reservoir
before returning to the system.
We speculate from the result of $N=2$ that, for general $N$, 
similar $n$-body bound states, $(1<n\leq N)$ 
contribute to the term of order $L^{-n}$ in the expectation value.
We assume that the contribution from the two-body
bound state is given by the function $I_{\pm}(k,h)$ 
in Eqs.~\eqref{eq:current-QM}.
Let $|k\rangle$
be an $N$-electron scattering state with
an incoming $N_{\alpha}$-electron plane wave
characterized by distinct wave-numbers $\{k_{i}^{\alpha}\}$ 
in the lead $\alpha$.
The speculated form of the expectation value is
\begin{align}
\frac{\langle k|I|k\rangle}{\langle k|k\rangle}
=&\frac{2\pi}{L}
  \Big(\sum_{i=1}^{N_{1}}
  I_{0}(k_{i}^{1})
  \!-\!\sum_{i=1}^{N_{2}}
  I_{0}(k_{i}^{2})
  \Big)
 \nn\\
&\!+\!\frac{4\pi^{2}}{L^{2}}
 \Big(
  \sum_{i<j}I_{+}(k_{i}^{1},\!k_{j}^{1})
  \!+\!\sum_{i,j}I_{-}(k_{i}^{1},\!k_{j}^{2})
  \!-\!\sum_{i<j}I_{+}(k_{i}^{2},\!k_{j}^{2})
 \Big)
 \!+\!O\Big(\frac{1}{L^{3}}\Big).
 \nn
\end{align}
We have verified this for $N=3$.
We neglect the terms of order higher than $L^{-2}$
in the expansion~\cite{Dhar-Sen-Roy_08PRL}.
By taking the limit $L, N_{\alpha}\to\infty$, 
the sum $(2\pi/L)\sum_{i=1}^{N_{\alpha}}$
should be replaced by the integral on $k$ 
with the zero-temperature
Fermi distribution $f_{\alpha}(k)=\theta(\mu_{\alpha}\!-\!k)$.
For $\mu_{1/2}\!=\!\pm V/2$, 
the average current is then given by
\begin{align}
\langle I\rangle
&=\int_{-V/2}^{V/2}\hspace{-10pt} dk\, I_{0}(k)
 +\frac{1}{2}\Big(\int_{-\Lambda}^{V/2}\hspace{-12pt} dk
  \int_{-\Lambda}^{V/2}\hspace{-12pt} dh\, 
 -\int_{-\Lambda}^{-V/2}\hspace{-12pt} dk
  \int_{-\Lambda}^{-V/2}\hspace{-12pt} dh\,
  \Big)
  I_{+}(k,h)
 \nn\\
&\quad
 +\!\int_{-\Lambda}^{V/2}\hspace{-12pt} dk
  \int_{-\Lambda}^{-V/2}\hspace{-15pt} dh\, 
  I_{-}(k,h),
\end{align}
where $-\Lambda$ is the low-energy cut-off. We have
\begin{align}
&\langle I\rangle
 \!=\!\frac{t^{2}}{2\pi}j_{-}
 \!+\!\frac{t^{2}}{8\pi^{2}}\frac{4U}{4+U^{2}}
  \Big(J\!-\!\frac{U}{2}J^{\prime}\Big),
\label{eq:noneq-current} \\
&J
 \!=\!2(\Bar{\Lambda}\!+\!j_{+})j_{2}
  \!+\!(j_{-}\!-\!j_{1})
  \log\frac{(\epsilon_{\Lambda}^{2}\!+\!1)^{2}}
  {(\epsilon_{+}^{2}\!+\!1)(\epsilon_{-}^{2}\!+\!1)},
 \nn\\
&J^{\prime}
 \!=\!2(\Bar{\Lambda}\!+\!j_{+})j_{1}
  \!+\!\Big(j_{2}+\frac{1}{2}
  \log\frac{\epsilon_{+}^{2}\!+1}{\epsilon_{-}^{2}\!+1}\Big)
  \log\frac{(\epsilon_{\Lambda}^{2}\!+\!1)^{2}}
  {(\epsilon_{+}^{2}\!+\!1)(\epsilon_{-}^{2}\!+\!1)},
 \nn
\end{align}
where $\epsilon_{\pm}=(\epsilon_{d}\pm V/2)/\Bar{t}^{2}$,
$\epsilon_{\Lambda}=(\epsilon_{d}+\Lambda)/\Bar{t}^{2}$,
$\Bar{\Lambda}=2(\Lambda/\Bar{t}^{2}-\arctan(\epsilon_{\Lambda}))$,
$j_{\pm}=\arctan(\epsilon_{+})\pm\arctan(\epsilon_{-})$ and
$j_{s}=\epsilon_{+}^{2-s}/(\epsilon_{+}^{2}+1)
 -\epsilon_{-}^{2-s}/(\epsilon_{-}^{2}+1)$, $(s=1,2)$.
The current includes higher-order terms in $U$
and, at $\epsilon_{d}=0$,
agrees with the perturbative result~\cite{Golub_07PRB} 
in the first order in $U$.
The linear divergence in $\Lambda\to\infty$ is due to 
the linearized dispersion relation in Eq.~\eqref{eq:IRLM_12}.
In Fig.~\ref{fig:current},
we plot the current-voltage characteristics
at $\epsilon_{d}=0$ by setting $\Lambda=V$.
The regime of negative differential conductance appears 
for large $U$~\cite{Doyon_07PRL,Boulat-Saleur-Schmitteckert_08PRL}. 

In summary, through the Landauer formula,
we have studied the nonequilibrium current
in an open quantum dot system by using exact scattering eigenstates.
We have found that the effect of the interaction appears
through the many-body bound states.
By taking the two-body bound state into account,
we have calculated the average current, which agrees with 
the perturbative result~\cite{Golub_07PRB} at $\epsilon_{d}=0$
and has a behavior similar to the other results~\cite{%
Doyon_07PRL,Boulat-Saleur-Schmitteckert_08PRL}.
In order to compare our result,
including the case $\epsilon_{d}\neq 0$, 
with the result in Ref.~\cite{Golub_07PRB} precisely, 
we need to consider contributions from other 
many-body bound states in Eq.~\eqref{eq:noneq-current},
because they may include first-order terms of $U$.
They would enable us to regularize
the logarithmic divergences in Eq.~\eqref{eq:noneq-current} with the
renormalization-group technique~\cite{Doyon-Andrei_06PRB,Doyon_07PRL}.

The authors would like to thank Dr.~T.~Fujii for discussions.
One of the authors (A.N.) also would like to thank
Prof.~T.~Deguchi for helpful comments.
The present study is partially supported by 
Grant-in-Aid for Young Scientists (B) No.~20740217,
Grant-in-Aid for Scientific Research (B) No.~17340115,
and CREST, JST.

\end{document}